\documentclass[aps,prb,reprint,twocolumn,superscriptaddress,longbibliography]{revtex4-2}
\pdfoutput=1

\usepackage{graphicx}
\usepackage{amsmath,amsfonts,amssymb}
\usepackage{bbold}
\usepackage{physics}
\usepackage{color}
\usepackage{hyperref}
\usepackage[normalem]{ulem}

\DeclareMathOperator{\diag}{diag}

\begin{document}

\title{Inherent momentum-dependent gap structure of altermagnetic superconductors}

\author{Christian L. H. Rasmussen}
\affiliation{Niels Bohr Institute, University of Copenhagen, DK-2200 Copenhagen, Denmark}

\author{Jannik Gondolf}
\affiliation{Niels Bohr Institute, University of Copenhagen, DK-2200 Copenhagen, Denmark}

\author{Mats Barkman}
\affiliation{Niels Bohr Institute, University of Copenhagen, DK-2200 Copenhagen, Denmark}

\author{Mercè Roig}
\affiliation{Department of Physics, University of Wisconsin–Milwaukee, Milwaukee, Wisconsin 53201, USA} 

\author{Daniel F. Agterberg}
\affiliation{Department of Physics, University of Wisconsin–Milwaukee, Milwaukee, Wisconsin 53201, USA} 

\author{Andreas Kreisel}
\affiliation{Niels Bohr Institute, University of Copenhagen, DK-2200 Copenhagen, Denmark}
\affiliation{Department of Physics and Astronomy, Uppsala University, Box 524, 751 20 Uppsala, Sweden}

\author{Brian M. Andersen}
\affiliation{Niels Bohr Institute, University of Copenhagen, DK-2200 Copenhagen, Denmark}

\date{May 14, 2026}

\begin{abstract}
Altermagnetic metals break time-reversal symmetry and feature spin-split Fermi surfaces generated by compensated Néel-ordered collinear magnetic moments. 
Being metallic, such altermagnets may undergo a further instability at low temperatures to a superconducting state, and it is an interesting open question what the salient features are of such altermagnetic superconductors.
We address this question on the basis of realistic microscopic models that capture the altermagnetic sublattice degrees of freedom.
We find that the sublattice structure can strongly affect the superconducting gap structure in altermagnetic superconductors.
In particular, it imposes nodes in the gap on the Brillouin zone edges for superconductors stabilized by momentum-independent bare attraction channels.
We contrast this to the case of superconductivity generated by extended range interactions where pairing is allowed on the Brillouin zone edges and both spin-singlet and equal-spin-pairing triplet
states can be stabilized.
Equal-spin-pairing triplet superconductivity is generically favored in the limit of large altermagnetic spin splitting of the bands compared to the superconducting gap scale, and features characteristic nonunitary properties arising from the altermagnetic order.
\end{abstract}

\maketitle

\section{Introduction}

Altermagnets comprise a separate class of collinear compensated magnetic materials in addition to conventional antiferromagnetism~\cite{Hayami2019,Ahn2019,Smejkal2020Jun,Yuan2020Jul,Mazin2021,Smejkal_Emerging_2022,Smejkal_Conventional_2022,Bhowal2022Dec,Reimers2024,Ding2024,Krempasky_Altermagnetic_2024,LeeMnTe2024,Osumi2024,Osumi2024,Fedchenko}. Distinct from standard antiferromagnets, the Néel order in altermagnets lacks the combined symmetry of time-reversal and inversion or translation, and features instead time-reversal and a rotation as a combined symmetry of the magnetically ordered state. This symmetry characteristic of altermagnets has important implications for their electronic band structure, most notably the existence of large momentum-dependent spin-split electronic bands, even in the absence of any net magnetization and relativistic spin-orbit coupling. In contrast, standard antiferromagnets without net magnetization preserve Kramers degeneracy and exhibit spin-degenerate bands throughout the Brillouin zone.

An important large subclass of altermagnets are altermagnetic metals. Such materials feature spin-split Fermi surfaces with associated spin-polarized low-energy quasiparticles potentially useful for spintronics devices~\cite{Smejkal_Emerging_2022}. Altermagnetic metals may also, however, undergo additional instabilities at lower temperatures, for example in the particle-particle channel stabilizing a superconducting phase. The resulting system is a new platform for studying the interesting interplay between magnetism and superconductivity. The main question is what unique properties the superconductor inherits when it emerges from an altermagnetic metal. 

Several previous theoretical works have investigated characteristic aspects of intrinsic superconductivity of altermagnets. For example, some studies have focused on consequences of the spin-split Fermi surfaces and its resulting unfavorable conditions for homogeneous spin-singlet pairing, leading instead to Fulde-Ferrell-Larkin-Ovchinnikov (FFLO) finite-momentum Cooper pairing, even in the absence of an external magnetic field~\cite{SotoGarrido2014,Chakraborty_Zerofield_2023,Sumita2023,Hong2025,Sim2024,HuFFLO,Mukasa2025,Iorsh2025,Sumita2025,Liu2025}. The resulting superconducting phase exhibits several unusual predicted properties, including e.g. Bogoliubov Fermi surfaces and the superconducting diode effect~\cite{Hong2025,Banerjee2024,Sim2024,Chakraborty_Diode,Cheng2024,RoigDiode}. Other theoretical studies have focused on the stabilization of unconventional superconducting phases as a consequence of the altermagnetic spin splitting~\cite{Bose2024,Hong2025,Brekke_Twodimensional_2023,Carvalho2024,Leraand2025,Parthenios,khodas2025tuningaltermagnetismstrain,Parshukov2025m,Monkman}. More specifically, several works have highlighted equal-spin-pairing (ESP) $p$-wave triplet states as the preferred superconducting order in the limit of large altermagnetic spin-split bands~\cite{Zhu2023,Hong2025,Franztorus,Monkman}. In this limit, the low-energy altermagnetic band structure essentially consists of distinct spin-polarized Fermi sheets, which naturally prefers ESP triplet superconductivity. The altermagnetic phase and its associated breaking of time-reversal symmetry may also be favorable for engineered topological superconductivity~\cite{Li2023Majoranacorner,Zhu2023,Li2024,Ghorashi2024}. Finally, we note a related research branch which, as opposed to the intrinsic altermagnetic superconductivity discussed above, focuses on the unique electronic properties from the proximity effect and hybrid systems between altermagnets and superconductors~\cite{Ouassou2023,Beenakker2023,Papaj2023,Zhang2024,Fukaya2025,Heinsdorf_proxy}. All these theoretical works are a testimony to the rich physics arising from the interplay between altermagnetic and superconducting instabilities.

Most of the above theory works have applied simplified one-band models to address the properties of altermagnetic superconductivity. Typically, the Hamiltonian is based on a low-energy model expanded around the center of the Brillouin zone with an imposed momentum-dependent Zeeman spin-splitting field. However, more realistic models necessarily include several magnetic sites per unit cell, in agreement with the fact that altermagnetic materials host a $q=0$ Néel order that does not expand the unit cell~\cite{Roig_Minimal_2024,Roig_Landau}. It remains an important question, what are the resulting superconducting properties of altermagnetic superconductors derived from minimal models that encompass the sublattice degree of freedom? Some recent works have addressed this question and e.g. explored constraints from the sublattice on the pairing structure~\cite{Chakraborty2024}, investigated sublattice effects favorable for stabilizing the phase-modulated FF phase (as opposed to LO)~\cite{Sumita2025}, and studied the consequences of the sublattice on characteristics of impurity-bound states in the superconducting phase~\cite{Jannik2025}. Recently, the key role of the sublattice degree of freedom for Cooper pairing was also highlighted in a study of superconductivity mediated by altermagnetic fluctuations~\cite{Wu2025}. Finally, we note that in the two-dimensional (2D) case, there exist altermagnets featuring no spin-splitting of the bands, yet they support an anomalous Hall effect~\cite{Jannik2025,Bai2025}. For the description of such systems, dubbed type IV altermagnets in Refs.~\cite{Bai2025,tian2025}, one-band models are not applicable and any proper theory must depart from models that include the relevant sublattice degree of freedom.

Here, we address the crucial role of the sublattice degree of freedom on pairing by combining recently developed minimal models for altermagnets and attractive density-density interactions capable of stabilizing superconductivity~\cite{Roig_Minimal_2024}. We restrict the discussion to $q=0$ superconductivity, which dominates the phase diagram of altermagnetic superconductivity~\cite{Chakraborty_Zerofield_2023,Hong2025,HuFFLO}. The main findings of our study include an inherent momentum dependence of the superconducting gap structure and unique altermagnetic fingerprints on equal-spin-pairing triplet superconductivity. The former point is most easily demonstrated in the case of superconductivity generated by featureless onsite attraction. In that case, altermagnets with Fermi surfaces close to the Brillouin zone (BZ) edges exhibit superconducting gap structures with significant momentum dependence. This is particularly prominent for Fermi surfaces crossing the BZ edges, where nodes in the pairing gap can arise even for onsite bare BCS-type attraction as relevant for phonon-mediated pairing. We stress that altermagnetism often is found in nonsymmorphic space groups where band degeneracies are guaranteed to be exhibited on the BZ edges. Since altermagnetism itself is favorable when interband contributions are important~\cite{Roig_Minimal_2024}, one expects on general grounds that the Fermi surfaces often will cross the BZ edges in these materials. Therefore, we expect the results for the superconducting gap anisotropy to be an important property of altermagnetic superconductors. In the case of superconductivity stabilized by extended range attractive interactions, both spin-singlet and spin-triplet superconducting order can be stabilized. Generally, for large spin-split bands compared to the superconducting gap scale ESP triplet order is preferred. Interestingly, we find that such ESP states are generally non-unitary and feature distinct signatures of the altermagnetic spin-splitting imprinted in their momentum-dependence. In addition, the transition to fully-gapped EPS triplet ground states generally takes place via two thermodynamic phase transitions.

\section{Methodology}\label{sec:methods}

\begin{figure*}[tb]
    \centering
    \includegraphics[width=\linewidth]{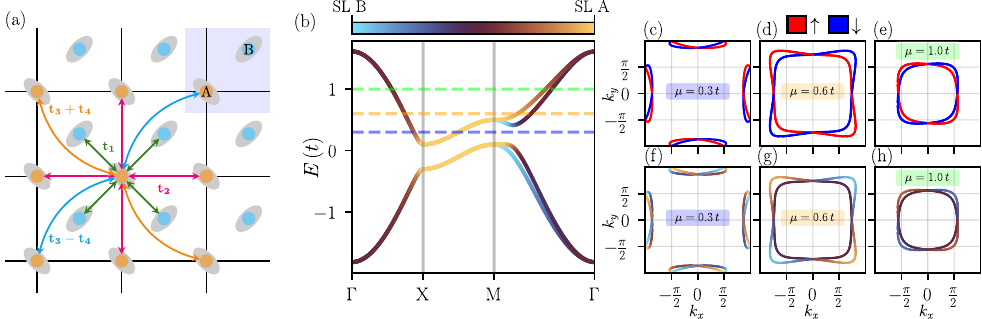}
    \caption{
    (a) Sublattice structure of the 2D lattice model with orange sites defining the $A$-sublattice and blue sites the $B$-sublattice. The different hoppings entering the model are indicated and labeled by different colors. (b) The electronic band structure along the high-symmetry path $\Gamma$-$X$-$M$-$\Gamma$ with sublattice coloring and three different Fermi levels indicated by the dashed horizontal lines. All energies are calculated in the energy scale $t$. Panels (c)-(h) show the Fermi surfaces for the three Fermi levels in (b). Panels (c)-(e) are colored according to spin and (f)-(h) according to sublattice weight.}
    \label{fig:model}
\end{figure*}
In this work, we apply minimal model Hamiltonians derived for centrosymmetric space groups with the magnetic atoms occupying inversion symmetric Wyckoff positions with multiplicity two~\cite{Roig_Minimal_2024}.
For simplicity, we restrict the study to the two-dimensional (2D) square lattice case with the two sublattice sites A and B separated by the half-lattice vector (1/2, 1/2). Thus, the applied minimal model of the altermagnetic metal is given by
\begin{align}
    H = \sum_{\mathbf{k}} c^\dagger_\mathbf{k} \mathcal{H}_\mathrm{MM} c_\mathbf{k},
\end{align}
with $c^\dagger_\mathbf{k} (c_\mathbf{k})$ denoting the creation (annihilation) operator written in the basis $c^\dagger_\mathbf{k} = (c^\dagger_{\uparrow A,\mathbf{k}},c^\dagger_{\uparrow B,\mathbf{k}}, c^\dagger_{\downarrow A,\mathbf{k}}, c^\dagger_{\downarrow B,\mathbf{k}})$ and  
\begin{equation}
    \mathcal{H}_\mathrm{MM}=\varepsilon_{0,\mathbf{k}} \tau_0 +t_{x,\mathbf{k}} \tau_x + t_{z,\mathbf{k}} \tau_z  + \tau_y {\boldsymbol \lambda}_\mathbf{k}\cdot {\boldsymbol \sigma} + \tau_z \mathbf{N} \cdot {\boldsymbol \sigma} ,
    \label{MMM}
\end{equation}
where $\sigma_i$ and $\tau_i$ are the Pauli matrices for the spin and sublattice degrees of freedom, respectively~\cite{Roig_Minimal_2024}. The particular momentum-dependent forms of the inter- ($t_{x,\mathbf{k}}$) and intra-sublattice ($t_{z,\mathbf{k}}$) site hopping terms and the spin-orbit coupling ${\boldsymbol \lambda}_\mathbf{k}$ are restricted by the properties of the layer group under consideration. The primary order parameter is the Néel order $\bf{N}$, and the resulting spin splitting of the altermagnetic state is driven by the development of a secondary magnetic octupole order in the Néel state, which in the minimal model is directly proportional to $t_{z,\mathbf{k}}$. The asymmetric hopping represented by $t_{z,\mathbf{k}}$ describes the local rotational symmetry breaking of the normal state, and transforms as a nontrivial irreducible representation (irrep) of the crystal point group. Thus, from the specific form of $t_{z,\mathbf{k}}$, it is possible to examine different types of altermagnetic phases. In Tab.~\ref{tab:tz_overview} we provide an overview of the form of $t_{z,\mathbf{k}}$, the associated altermagnetic spin splitting, and their relevance for exemplary layer groups. We include the case of type IV altermagnets with spin-degenerate bands in the non-relativistic limit~\cite{Jannik2025,Bai2025,tian2025}.

\begin{table}[b]
\caption{Overview of different finite altermagnetic spin-splitting terms for 2D layer groups. We show the form of $t_{z,\mathbf{k}}$, the associated momentum-dependent spin splitting, and exemplary layer groups for which it is relevant, see also Ref.~\cite{Jannik2025}. The last row refers to type IV altermagnets, featuring no spin splitting in the band structure.}
\centering
\begin{ruledtabular}
\begin{tabular}{ccc}
$t_{z,\mathbf{k}}$ & Nodal-structure & Layer groups \\
\hline
$\cos{k_x}-\cos{k_y}$ & $d_{x^2-y^2}$ & L61 \\
$\sin{k_x}\sin{k_y}$ & $d_{xy}$ & L17, L44, L63 \\
$\sin{k_x}\sin{k_y}(\cos{k_x}-\cos{k_y})$ & $g$ & L63 \\
0 & no spin splitting & L42 \\
\end{tabular}
\end{ruledtabular}
\label{tab:tz_overview}
\end{table}

In the following, for concreteness we focus initially on altermagnets with the nodal structure of $d_{xy}$-form generated by $t_{z,\mathbf{k}}= 4t_4 \sin{k_x}\sin{k_y}$, and discuss the other cases further below. The sublattice-independent dispersion is given by $\varepsilon_{0,\mathbf{k}} = -2t_2\left(\cos{k_x}+\cos{k_y}\right) -4 t_3\left(\cos{k_x}\cos{k_y}\right) - \mu$ whereas the inter-sublattice hopping is given by $t_{x,\mathbf{k}} = -4t_1 \cos{\frac{k_x}{2}}\cos{\frac{k_y}{2}}$. The hopping parameters are similar to those found in Ref.~\cite{Jannik2025}, and are set to $\{t_1,t_2,t_3,t_4\} = \{0.425,0.05,-0.025,-0.075\} t$, where $t$ is the energy scale used throughout this work.

In the following, we neglect spin-orbit coupling (SOC) given by the term $\mathbf{\lambda}_\mathbf{k}$, and choose the quantization axis of the Néel order parameter $\mathbf{N}=N\mathbf{z}$ along the $z$ direction. We set the Néel order parameter to $N=0.2$ throughout the results section. Similar to Ref.~\cite{Jannik2025}, we perform a gauge transformation of the B sublattice creation and annihilation operators as $c_{\sigma B,\mathbf{k}}\rightarrow c_{\sigma B,\mathbf{k}} e^{i \phi_\mathbf{k}}$. In the new basis, $t_{x,\mathbf{k}}\rightarrow t_{x,\mathbf{k}}e^{i \phi_\mathbf{k}}$ and the resulting Hamiltonian matrix exhibits the full periodicity of the BZ for $\phi_\mathbf{k}=-\frac{k_x}{2}-\frac{k_y}{2}$. The minimal model is thus given by
\begin{equation}\label{eq:minimalModel}
    \mathcal{H} = \varepsilon_{0,\mathbf{k}} \tau_0 + t_{z,\mathbf{k}} \tau_z + \Re[t_{x,\mathbf{k}}] \tau_x - \Im[t_{x,\mathbf{k}}] \tau_y + N \tau_z \sigma_z.
\end{equation}
The matrix $\mathcal{H}$ is block-diagonal in the spin index $\sigma$ and each block can be diagonalized as $U_{\sigma, \mathbf{k}}^\dagger \mathcal{H}_\sigma U_{\sigma, \mathbf{k}} = \diag(E^{\alpha}_{\sigma,\mathbf{k}},E^{\beta}_{\sigma,\mathbf{k}})$ with eigenvalues $E^\lambda_{\sigma,\mathbf{k}}= \varepsilon_{0,\mathbf{k}} \pm \sqrt{|t_{x,\mathbf{k}}|^2 + (t_{z,\mathbf{k}} + \sigma N)^2}$ ($+$ for the upper band $\alpha$ and $-$ for the lower band $\beta$) and unitary transformation
\begin{equation}\label{eq:unitaryTrafo}
    U_{\sigma,\mathbf{k}} =
        \begin{pmatrix}
            \cos{\frac{\theta_\mathbf{\sigma,k}}{2}} & \frac{t_x}{\abs{t_x}}\sin{\frac{\theta_\mathbf{\sigma,k}}{2}} \\
            -\frac{t_x^*}{\abs{t_x}} \sin{\frac{\theta_\mathbf{\sigma,k}}{2}} & \cos{\frac{\theta_\mathbf{\sigma,k}}{2}}
        \end{pmatrix},
\end{equation}
where
\begin{align}
    \cos{\frac{\theta_\mathbf{\sigma,k}}{2}} &= \frac{1}{\sqrt{2}}\sqrt{1+\frac{t_{z,\mathbf{k}}+\sigma N}{\sqrt{|t_{x,\mathbf{k}}|^2 + (t_{z,\mathbf{k}} + \sigma  N)^2}}},\\
    \sin{\frac{\theta_\mathbf{\sigma,k}}{2}} &= \frac{-1}{\sqrt{2}}\sqrt{1-\frac{t_{z,\mathbf{k}}+\sigma N}{\sqrt{|t_{x,\mathbf{k}}|^2 + (t_{z,\mathbf{k}} + \sigma N)^2}}}. 
\end{align}

Figure~\ref{fig:model} shows an exemplary $d$-wave altermagnet for layer group L63. The anisotropic crystal environments and hoppings for the sublattices are shown in Fig.~\ref{fig:model}(a). The spin-split Fermi surface and band structure can be seen in Fig.~\ref{fig:model}(b-h), also showing the mixing of the sublattice and pure spin character of the bands. The primary distinction of the minimal model Eq.~\eqref{MMM} from one-band models is the incorporated sublattice weights defined through the functions,
    \begin{align}
    l_{\mathbf{\sigma,k}} &= \cos{\frac{\theta_\mathbf{\sigma,k}}{2}}, \\
    m_{\mathbf{\sigma,k}} &= \frac{t_x}{\abs{t_x}}\sin{\frac{\theta_\mathbf{\sigma,k}}{2}}, \label{eq:m} 
\end{align}
where $ l_{\mathbf{\sigma,k}}$ and $ m_{\mathbf{\sigma,k}}$ are the unitary transformation components between sublattice and band space. The particular form of Eq.~(\ref{eq:m}) ensures $2\pi$ periodicity in the BZ of the unitary transformation,
\begin{align}
    \begin{pmatrix}
        d_\mathbf{\sigma,\alpha,k} \\
        d_\mathbf{\sigma,\beta,k}
    \end{pmatrix} = 
    \begin{pmatrix}
        l_{\mathbf{\sigma,k}} & m_{\mathbf{\sigma,k}} \\
        -m^\ast_{\mathbf{\sigma,k}} & l_{\mathbf{\sigma,k}} 
    \end{pmatrix} \begin{pmatrix}
        c_{\sigma,A,\mathbf{k}} \\
        c_{\sigma,B,\mathbf{k}}
    \end{pmatrix}.
\end{align}

In the band basis, the Hamiltonian reads
\begin{equation}
    H = \sum_{\mathbf{k}, \sigma} \Bigl[E^\alpha_{\sigma, \mathbf{k}} d^\dagger_{\sigma,\alpha, \mathbf{k}}d_{\sigma,\alpha, \mathbf{k}} +
    E^\beta_{\sigma, \mathbf{k}} d^\dagger_{\sigma,\beta, \mathbf{k}}d_{\sigma,\beta,\mathbf{k}}\Bigr].
\end{equation}
To stabilize superconductivity, we consider the general interaction,  
\begin{align}
    H_{\mathrm{int}} &= 
    -\sum_{i,j,s,s'} V^{s,s'}_{i,j} c^\dagger_{\sigma,s,\mathbf r_i}c^\dagger_{\sigma',s',\mathbf r_j}c_{\sigma',s',\mathbf r_j}c_{\sigma,s,\mathbf r_i}, \label{H_int}
\end{align}
where $\mathbf r_i$ and $\mathbf r_j$ denote the positions of the unit cell of sublattice $s$ and $s'$. In the following, we consider both onsite (OS) and extended interactions, and discuss how the two-sublattice model differs from a one-band model. We write the interaction in momentum space, anticipating mean-field decoupling in the Cooper channel without finite-momentum pairing. The magnitude of the OS-interaction is identical on the two sublattice sites, such that the interaction reads
\begin{align}\label{eq:HOS}
    H^{\mathrm{OS}}_\mathrm{int} =-\frac{V}{N}\sum_{s,\mathbf{k,k'}} & c^\dagger_{\uparrow,s,\mathbf{k}}c^\dagger_{\downarrow,s,\mathbf{-k}}c_{\downarrow,s,\mathbf{-k'}}c_{\uparrow,s,\mathbf{k'}}, 
\end{align}
where $V > 0$ is the onsite attraction, $N$ is the system size and $s \in \{\mathrm{A,B}\}$ denotes the sublattice degree of freedom. 
For the extended interactions we consider both nearest-neighbor (NN) and next-nearest-neighbor (NNN) attraction given by the Hamiltonian
\begin{align}
    H^{\mathrm{ext}}_\mathrm{int} =-\frac{1}{N}
    \sum_{\sigma, \sigma'}
    \sum_{s,\mathbf{k,k'}} & V^\mathrm{ext}_{\mathbf{k,k'},s,s'} c^\dagger_{\sigma,s,\mathbf{k}}c^\dagger_{\sigma',s',\mathbf{-k}}c_{\sigma',s',\mathbf{-k'}}c_{\sigma,s,\mathbf{k'}}, \label{eq:H_int_NN}
\end{align}
where the interaction in the NN case connects opposite sublattices $s' = \Bar{s}$, whereas the interaction for the NNN case is between same sublattices $s'=s$. For the case of NN attraction the bare potential is given by
\begin{align}
V^{\mathrm{NN}}_{\mathbf{k,k'},\mathrm{A,B}}& =  4V^{\mathrm{NN}}e^{-i\frac{k_x+k_y}{2}}e^{i\frac{k'_x+k'_y}{2}}
 \nonumber\\&\times
\cos{\left(\frac{k_x-k_x'}{2}\right)} \cos{\left(\frac{k_y-k_y'}{2}\right)},
\end{align}
where $V^\mathrm{NN}>0$ denotes the strength of the NN attraction. Note that the phase associated with the NN interaction is dependent on the sublattice such that $V^{\mathrm{NN}}_{\mathbf{k,k'},\mathrm{B,A}}={V^{\mathrm{NN}}_{\mathbf{k,k'},\mathrm{A,B}}}^*$. The phase factor arises from the specific choice of basis and only appears in interactions between different sublattices. Superconducting order parameters expressed in different bases are related by a unitary transformation, and all gauge-invariant observables remain invariant under such changes of basis.
For the case of NNN attraction, the bare potential is given by
\begin{align}
V^{\mathrm{NNN}}_{\mathbf{k,k'},\mathrm{s,s}}& = 2 V^{\mathrm{NNN}} 
\left(
\cos{\left(k_x - k_x'\right)}+\cos{\left(k_y - k_y'\right)}
\right),
\end{align}
where $V^\mathrm{NNN} > 0 $ denotes the strength of the NNN attraction.

The operators in Eq.~(\ref{eq:HOS}) and Eq.~(\ref{eq:H_int_NN}) are expressed as linear combinations of operators in band space using $l_{\mathbf{\sigma,k}}$ and $m_{\mathbf{\sigma,k}}$, resulting in sixteen terms that incorporate various combinations of the eigenvectors mentioned above. At the specified Fermi levels (see Fig.~\ref{fig:model}), the separation between the bands is substantial and the restriction to the intra-band terms is a good approximation. This is confirmed by an explicit comparison of the spectrum obtained with and without inter-band pairing, which shows only minor quantitative deviations along the $M$--$X$ path where the different bands are closest to each other, but no qualitative changes as long as the inter-band gap is smaller than the magnetic energy scale dictated by $N$.
In the following, we will first discuss the case of OS attraction. Upon restriction to the intra-band terms, the Hamiltonian Eq.~(\ref{eq:HOS}) reduces to the following form
\begin{widetext}
\begin{align}
    H^\mathrm{OS}_{\mathrm{int}} = -\frac{V}{N}\sum_{\mathbf{k,k'}}\Big[&\left(l_{\uparrow,\mathbf{k}}l_{\uparrow,\mathbf{k'}}l_{\downarrow,-\mathbf{k}}l_{\downarrow,\mathbf{-k'}} + m_{\uparrow,\mathbf{k}}m^\ast_{\uparrow,\mathbf{k'}}m_{\downarrow,-\mathbf{k}}m^\ast_{\downarrow,-\mathbf{k'}}\right) d^\dagger_{\uparrow,\alpha,\mathbf{k}}d_{\uparrow,\alpha,\mathbf{k'}}d^\dagger_{\downarrow,\alpha,\mathbf{-k}}d_{\downarrow,\alpha,-\mathbf{k'}} \nonumber \\
    +&\left(m^\ast_{\uparrow,\mathbf{k}}m_{\uparrow,\mathbf{k'}}m^\ast_{\downarrow,-\mathbf{k}}m_{\downarrow,-\mathbf{k'}}  +l_{\uparrow,\mathbf{k}}l_{\uparrow,\mathbf{k'}}l_{\downarrow,-\mathbf{k}}l_{\downarrow,-\mathbf{k'}} \right) d^\dagger_{\uparrow,\beta,\mathbf{k}}d_{\uparrow,\beta,\mathbf{k'}}d^\dagger_{\downarrow,\beta,\mathbf{-k}}d_{\downarrow,\beta,-\mathbf{k'}} \nonumber \\
    +&\left(l_{\uparrow,\mathbf{k}}m_{\uparrow,\mathbf{k'}}l_{\downarrow,-\mathbf{k}}m_{\downarrow,-\mathbf{k'}} + m_{\uparrow,\mathbf{k}}l_{\uparrow,\mathbf{k'}}m_{\downarrow,-\mathbf{k}}l_{\downarrow,-\mathbf{k'}}\right) d^\dagger_{\uparrow,\alpha,\mathbf{k}}d_{\uparrow,\beta,\mathbf{k'}}d^\dagger_{\downarrow,\alpha,\mathbf{-k}}d_{\downarrow
    ,\beta,-\mathbf{k'}} \nonumber \\
    +&\left(m^\ast_{\uparrow,\mathbf{k}}l_{\uparrow,\mathbf{k'}}m^\ast_{\downarrow,-\mathbf{k}}l_{\downarrow,-\mathbf{k'}} + l_{\uparrow,\mathbf{k}}m^\ast_{\uparrow,\mathbf{k'}}l_{\downarrow,-\mathbf{k}}m^\ast_{\downarrow,-\mathbf{k'}}\right) d^\dagger_{\uparrow,\beta,\mathbf{k}}d_{\uparrow,\alpha,\mathbf{k'}}d^\dagger_{\downarrow,\beta,\mathbf{-k}}d_{\downarrow
    ,\alpha,-\mathbf{k'}} \Big] + \mbox{H.c.}, \label{eq:H_int_band}
\end{align}
\end{widetext}
where $d_{\sigma,\alpha,\mathbf{k}}$ and $d_{\sigma,\beta,\mathbf{k}}$ denote the different fermionic operators for each band.
The above equation only allows for pairwise combinations of the eigenvectors, such that the interaction always exhibits the combinations $l_{\uparrow,\mathbf{k}}l_{\downarrow,\mathbf{-k}}$ and $m_{\uparrow,\mathbf{k}}m_{\downarrow,\mathbf{-k}}$.  In order to examine the superconducting order parameter, we employ the self-consistent mean-field Bogoliubov-de-Gennes (BdG) approach. Consequently, the Hamiltonian is decoupled under the mean-field approximation, resulting in the Nambu matrix
\begin{align}
   \mathcal{H}^\mathrm{OS}_\mathrm{BdG} &= \begin{pmatrix}
       \hat{E}_\mathbf{k} & \hat{\Delta}_\mathbf{k} \\ \hat{\Delta}^\ast_\mathbf{k} & -\hat{E}_\mathbf{-k} 
   \end{pmatrix}, \label{eq:H_BdG}
\end{align} 

\begin{align} 
   \hat{E}_\mathbf{k} &= \begin{pmatrix}
       \hat{E}^\alpha_\mathbf{k} & 0 \\ 0 & \hat{E}^\beta_\mathbf{k} 
   \end{pmatrix},  &\quad \hat{\Delta}_\mathbf{k} &= \begin{pmatrix}
       \hat{\Delta}^\alpha_\mathbf{k} & 0 \\ 0 & \hat{\Delta}^\beta_\mathbf{k} 
   \end{pmatrix}, \\
   \hat{E}^\lambda_\mathbf{k} &= \begin{pmatrix}
       E^\lambda_{\uparrow,\mathbf{k}} & 0 \\ 0 & E^\lambda_{\downarrow,\mathbf{k}} 
   \end{pmatrix}, &\quad \hat{\Delta}^\lambda_\mathbf{k} &= \begin{pmatrix}
       0 & \Delta^\lambda_{\uparrow\downarrow,\mathbf{k}} \\ \Delta^\lambda_{\downarrow\uparrow,\mathbf{k}} & 0   \end{pmatrix},
\end{align}
with the eigenenergies on the diagonal and $\lambda \in \{\alpha,\beta\}$. The superconducting order parameter from this approach is given by,
\begin{align}
    &\Delta^{\alpha}_{\uparrow\downarrow,\mathbf{k}} = -\frac{V}{N} \Big[l_{\uparrow,\mathbf{k}}l_{\downarrow,\mathbf{-k}} \sum_{\mathbf{k'}}l_{\uparrow,\mathbf{k'}}l_{\downarrow,\mathbf{-k'}}\langle d_{\downarrow,\alpha,-\mathbf{k'}}d_{\uparrow,\alpha,\mathbf{k'}}\rangle\nonumber \\  &+ m_{\uparrow,\mathbf{k}}m_{\downarrow,\mathbf{-k}}\sum_{\mathbf{k'}}m^\ast_{\uparrow,\mathbf{k'}}m^\ast_{\downarrow,\mathbf{-k'}}\langle d_{\downarrow,\alpha,-\mathbf{k'}}d_{\uparrow,\alpha,\mathbf{k'}}\rangle  \nonumber \\
    &  + l_{\uparrow,\mathbf{k}}l_{\downarrow,\mathbf{-k}} \sum_{\mathbf{k'}}m_{\uparrow,\mathbf{k'}}m_{\downarrow,\mathbf{-k'}} \langle d_{\downarrow,\beta,-\mathbf{k'}}d_{\uparrow,\beta,\mathbf{k'}} \rangle \nonumber \\
    &+ m_{\uparrow,\mathbf{k}}m_{\downarrow,\mathbf{-k}}\sum_{\mathbf{k'}}l_{\uparrow,\mathbf{k'}}l_{\downarrow,\mathbf{-k'}}\langle d_{\downarrow,\beta,-\mathbf{k'}}d_{\uparrow,\beta,\mathbf{k'}} \rangle\Big], \label{eq:SC_OP_A}
\end{align}
where the superconducting order parameter for band $\alpha$ has contributions from both bands. As seen from Eq.~(\ref{eq:SC_OP_A}), $\Delta^{\alpha}_{\uparrow\downarrow,\mathbf{k}}$ clearly acquires momentum dependence through the eigenvectors for the case of onsite bare attractions. The gap $\Delta^\beta_{\uparrow\downarrow,\mathbf{k}}$, which also enters the BdG Hamiltonian in Eq.~(\ref{eq:H_BdG}), can likewise be expressed as above, but in the weak-coupling limit its value has limited physical relevance since the $\beta$ band does not cross the Fermi level. We therefore do not discuss $\Delta^\beta$ in the following. Since the interaction creates off-diagonal elements, we diagonalize it by a Bogoliubov transformation with the eigenenergies written in terms of the average and the difference of spin-split energies,
\begin{align}
    E^{\lambda,\pm}_{\sigma\Bar{\sigma},\mathbf{k}} &= \delta E^\lambda_{\sigma\Bar{\sigma},\mathbf{k}} \pm E^{\lambda,0}_{\sigma\Bar{\sigma},\mathbf{k}}, \\
    E^{\lambda,0}_{\sigma\Bar{\sigma},\mathbf{k}} &= \sqrt{(\Bar{\varepsilon}^\lambda_{\mathbf{k}})^2+|\Delta^\lambda_{\sigma\Bar{\sigma},\mathbf{k}}|^2}, \\
    \delta E^\lambda_{\sigma\Bar{\sigma},\mathbf{k}} &= \frac{E^\lambda_{\sigma,\mathbf{k}} - E^\lambda_{\Bar{\sigma},\mathbf{k}}}{2}, \quad 
    \Bar{\varepsilon}^\lambda_{\mathbf{k}} = \frac{E^\lambda_{\sigma,\mathbf{k}} + E^\lambda_{\Bar{\sigma},\mathbf{k}}}{2},
\end{align}
incorporating the combinations $\sigma\Bar{\sigma}$ that represent configurations with opposite spin orientations. The Nambu eigenvector $(u^\lambda_{\sigma\Bar{\sigma},\mathbf{k}}, v^\lambda_{\sigma\Bar{\sigma}\mathbf{k}})^T$ equals
\begin{align}
     u^\lambda_{\sigma\Bar{\sigma},\mathbf{k}} &= \sqrt{\frac{1}{2}\left(1 + \frac{\Bar{\varepsilon}^\lambda_{\mathbf{k}}}{E^{\lambda,0}_{\sigma\Bar{\sigma},\mathbf{k}}}\right)}, \\
     v^\lambda_{\sigma\Bar{\sigma}\mathbf{k}} &= \frac{\Delta^{\lambda}_{\sigma\Bar{\sigma},\mathbf{k}}}{|\Delta^{\lambda}_{\sigma\Bar{\sigma},\mathbf{k}}|} \sqrt{\frac{1}{2}\left(1 - \frac{\Bar{\varepsilon}^\lambda_{\mathbf{k}}}{E^{\lambda,0}_{\sigma\Bar{\sigma},\mathbf{k}}}\right)}.
\end{align}
Hence, we express Eq.~(\ref{eq:SC_OP_A}) in terms of quasiparticle operators by using the above eigenvectors and calculate the superconducting order parameter self-consistently by iterating the following gap equation until convergence is reached,
\begin{align}
    \Delta^\alpha_{\uparrow\downarrow,\mathbf{k}} &= -\frac{V}{N}\Big[l_{\uparrow,\mathbf{k}}l_{\downarrow,\mathbf{-k}} \sum_{\mathbf{k'}}l_{\uparrow,\mathbf{k'}}l_{\downarrow,\mathbf{-k'}}\frac{\Delta^{\alpha\ast}_{\uparrow\downarrow,\mathbf{k'}}}{2E^{\alpha,0}_{\uparrow\downarrow,\mathbf{k'}}}(1-2f(E^{\alpha,+}_{\uparrow\downarrow,\mathbf{k'}})) \nonumber \\
    & + m_{\uparrow,\mathbf{k}}m_{\downarrow,\mathbf{-k}}\sum_{\mathbf{k'}} m_{\uparrow,\mathbf{k'}}^\ast m^\ast_{\downarrow,\mathbf{-k'}}\frac{\Delta^{\alpha\ast}_{\uparrow\downarrow,\mathbf{k'}}}{2E^{\alpha,0}_{\uparrow\downarrow,\mathbf{k'}}}(1-2f(E^{\alpha,+}_{\uparrow\downarrow,\mathbf{k'}}))\nonumber \\
    & + l_{\uparrow,\mathbf{k}}l_{\downarrow,\mathbf{-k}} \sum_{\mathbf{k'}}m_{\uparrow,\mathbf{k'}}m_{\downarrow,\mathbf{-k'}}\frac{\Delta^{\beta\ast}_{\uparrow\downarrow,\mathbf{k'}}}{2E^{\beta,0}_{\uparrow\downarrow,\mathbf{k'}}}(1-2f(E^{\beta,+}_{\uparrow\downarrow,\mathbf{k'}})) \nonumber \\
    &  +m_{\uparrow,\mathbf{k}}m_{\downarrow,\mathbf{-k}}\sum_{\mathbf{k'}} l_{\uparrow,\mathbf{k'}}l_{\downarrow,\mathbf{-k'}}\frac{\Delta^{\beta\ast}_{\uparrow\downarrow,\mathbf{k'}}}{2E^{\beta,0}_{\uparrow\downarrow,\mathbf{k'}}}(1-2f(E^{\beta,+}_{\uparrow\downarrow,\mathbf{k'}}))\Big], \label{eq:OP_A_BdG}
\end{align}
with $f(E^{\lambda,\pm}_{\sigma\Bar{\sigma},\mathbf{k}})$ denoting the Fermi-Dirac distribution function. We stress that the momentum structure in the onsite case stems solely from the eigenvectors from sublattice to band space, leading to significant changes from the simple one-band model.

We end this section by returning to the case where superconductivity is stabilized by extended range attractions as given by the BCS Hamiltonian in Eq.~(\ref{eq:H_int_NN}). For the case of NN attraction, the relevant sublattices are confined to the opposite sublattices, as these constitute the nearest neighbors. Hence, the interacting Hamiltonian restricted to the intra-band terms, $\mathcal{H}^\mathrm{NN}_{\mathrm{int}}$, acquires a different form due to different combinations of the eigenvectors. Calculating the superconducting order parameter from NN attraction, we obtain
\begin{align}
\Delta^{\alpha,\mathrm{NN}}_{\sigma\sigma',\mathbf{k}} &=\nonumber\\
-\frac 1N \Big[&l_{\sigma,\mathbf{k}}m_{\sigma',\mathbf{-k}}\sum_{\mathbf{k'}}V^{\mathrm{NN}}_{\mathbf{k,k'},\mathrm{A,B}}m^\ast_{\sigma',\mathbf{-k'}} l_{\sigma,\mathbf{k'}} \langle d_{\sigma',\alpha,-\mathbf{k'}}d_{\sigma,\alpha,\mathbf{k'}}\rangle\nonumber \\
+ &m_{\sigma,\mathbf{k}} l_{\sigma',-\mathbf{k}} \sum_{\mathbf{k'}} V^{\mathrm{NN}}_{\mathbf{k,k'},\mathrm{B,A}}l_{\sigma',-\mathbf{k'}} m^\ast_{\sigma,\mathbf{k'}} \langle d_{\sigma',\alpha,-\mathbf{k'}}d_{\sigma,\alpha,\mathbf{k'}}\rangle
\nonumber \\
- &l_{\sigma,\mathbf{k}}m_{\sigma',-\mathbf{k}}\sum_{\mathbf{k'}}V^{\mathrm{NN}}_{\mathbf{k,k'},\mathrm{A,B}}l_{\sigma',-\mathbf{k'}} m_{\sigma,\mathbf{k'}} \langle d_{\sigma',\beta,-\mathbf{k'}}d_{\sigma,\beta,\mathbf{k'}} \rangle \nonumber \\
- &m_{\sigma,\mathbf{k}} l_{\sigma',-\mathbf{k}} \sum_{\mathbf{k'}}V^{\mathrm{NN}}_{\mathbf{k,k'},\mathrm{B,A}}m_{\sigma',-\mathbf{k'}} l_{\sigma,\mathbf{k'}} \langle d_{\sigma',\beta,-\mathbf{k'}}d_{\sigma,\beta,\mathbf{k'}} \rangle\Big]. \label{eq:Delta_A_NN}
\end{align}

Likewise, for the case of superconductivity generated by NNN attractions, the resulting gap equation becomes,
\begin{align}
\Delta^{\alpha,\mathrm{NNN}}_{\sigma\sigma',\mathbf{k}}&= \nonumber \\
-\frac 1N \Big[&l_{\sigma,\mathbf{k}}l_{\sigma',\mathbf{-k}}\sum_{\mathbf{k'}}V^{\mathrm{NNN}}_{\mathbf{k,k'},\mathrm{A,A}}l^\ast_{\sigma',\mathbf{-k'}} l_{\sigma,\mathbf{k'}} \langle d_{\sigma',\alpha,-\mathbf{k'}}d_{\sigma,\alpha,\mathbf{k'}}\rangle\nonumber \\
+ &m_{\sigma,\mathbf{k}}m_{\sigma',-\mathbf{k}} \sum_{\mathbf{k'}} V^{\mathrm{NNN}}_{\mathbf{k,k'},\mathrm{A,A}}m^\ast_{\sigma',-\mathbf{k'}} m^\ast_{\sigma,\mathbf{k'}} \langle d_{\sigma',\alpha,-\mathbf{k'}}d_{\sigma,\alpha,\mathbf{k'}}\rangle
\nonumber \\
+&l_{\sigma,\mathbf{k}}l_{\sigma',-\mathbf{k}}\sum_{\mathbf{k'}}V^{\mathrm{NNN}}_{\mathbf{k,k'},\mathrm{B,B}}m_{\sigma',-\mathbf{k'}} m_{\sigma,\mathbf{k'}} \langle d_{\sigma',\beta,-\mathbf{k'}}d_{\sigma,\beta,\mathbf{k'}} \rangle \nonumber \\
+ &m_{\sigma,\mathbf{k}} m_{\sigma',-\mathbf{k}} \sum_{\mathbf{k'}}V^{\mathrm{NNN}}_{\mathbf{k,k'},\mathrm{B,B}}l_{\sigma',-\mathbf{k'}} l_{\sigma,\mathbf{k'}} \langle d_{\sigma',\beta,-\mathbf{k'}}d_{\sigma,\beta,\mathbf{k'}} \rangle\Big].
\label{eq:Delta_A_NNN}
\end{align}

The key difference arising from the sublattice combination is that we now have different products of $l_{\sigma,\mathbf{k}}$ and $m_{\sigma,\mathbf{k}}$ compared to the case with onsite interactions.
The $V^\mathrm{NN}_{\mathbf{k,k'},s,\bar{s}}$ and $V^\mathrm{NNN}_{\mathbf{k,k'},s,s}$ interactions can be decoupled through basic trigonometric identities to separate the momentum dependence in $\mathbf{k}$ and $\mathbf{k'}$. Having rewritten the interaction as product of functions of $\mathbf{k}$ and $\mathbf{k'}$, we compute the superconducting order parameter using the BdG transformation. Evaluating the expectation values, we arrive at a corresponding self-consistent gap equation expressed in terms of the eigenvalues and eigenvectors similar to Eq.~(\ref{eq:OP_A_BdG}).

\section{Results}\label{sec:results}

\subsection{Onsite attraction}
\begin{figure}[tb]
    \centering
    \includegraphics[width=\linewidth]{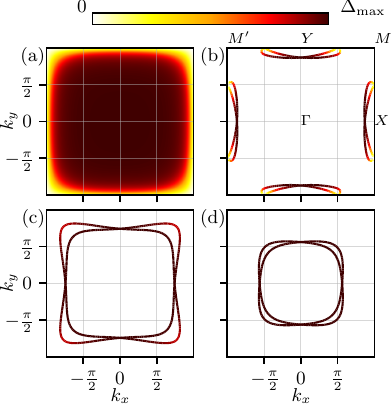}
    \caption{The superconducting order parameter stabilized by onsite bare attraction. (a) The structure of the combined form factor $l_{\uparrow,\mathbf{k}}l_{\downarrow,\mathbf{-k}}+m_{\uparrow,\mathbf{k}}m_{\downarrow,\mathbf{-k}}$. (b)-(d) display the superconducting gap structure on the Fermi surface for the chemical potentials shown in Fig.~\ref{fig:model}(b). The onsite interaction used for the results shown in (b)-(d) is $V=1.5t$. $\Delta_\mathrm{max}$ for (b) $0.17t$, (c) $0.24t$ and (d) $0.18t$. The superconducting order parameter is color-coded according to its maximum value. The coordinates to the path used in Fig.~\ref{fig:coeff_ll_mm_spin_weights}(c)-(d) is marked in (b).}
    \label{fig:gap_amplitude}
\end{figure}

In Fig.~\ref{fig:gap_amplitude} we show the superconducting gap structure obtained for the case of onsite bare attraction. Figure~\ref{fig:gap_amplitude}(a) displays the important form factor $l_{\uparrow,\mathbf{k}}l_{\downarrow,\mathbf{-k}}+m_{\uparrow,\mathbf{k}}m_{\downarrow,\mathbf{-k}}$ throughout the BZ, whereas Figs.~\ref{fig:gap_amplitude}(b-d) show the gap on the same three Fermi surfaces displayed in Fig.~\ref{fig:model}(c-h). As seen, significant momentum dependence is exhibited for Fermi surfaces reaching the BZ edge, arising from the sublattice-to-band matrix elements.

\begin{figure}
    \centering
    \includegraphics[width=\linewidth]{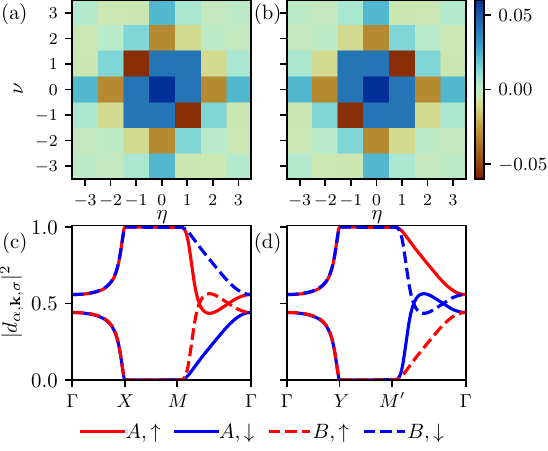}
    \caption{
    Fourier coefficients $c^i_{\nu,\eta}$ from the decoupling of (a) $l_{\uparrow,\mathbf{k}}l_{\downarrow,\mathbf{-k}}$ and (b) $m_{\uparrow,\mathbf{k}}m_{\downarrow,\mathbf{-k}}$. The color scale is chosen such that the structure of the neighboring sites are clearly visible, as the onsite value $c^l_{00} = c^m_{00} = 0.414$ dominates. The sublattice weights on the $d_{\alpha,\mathbf{k},\sigma}$ band with color-scheme red (blue) for spin-up (down) and solid (dashed) for sublattice A (B). (c) is along the path $\Gamma-X-M-\Gamma$ and (d) along the path $\Gamma-Y-M^\prime-\Gamma$ shown in Fig.~\ref{fig:gap_amplitude}(b)}
    \label{fig:coeff_ll_mm_spin_weights}
\end{figure}

In order to quantify the momentum dependence of the order parameter, we can perform a Fourier decomposition of the eigenvector combinations 
\begin{align}
    l_{\uparrow,\mathbf{k}}l_{\downarrow,\mathbf{-k}} = \sum_{\nu,\eta} c^l_{\nu,\eta} e^{-i k_x \nu} e^{-i k_y \eta},
\end{align}
where the coefficients $c^l_{\nu,\eta}$ denote the contribution to higher-order Fourier components.
The coefficient $c^l_{00}$ refers to the onsite component of the decoupling. A similar decomposition can be performed of the $m$ eigenvector combination, $m_{\uparrow,\mathbf{k}}m_{\downarrow,\mathbf{-k}}$, yielding the coefficients $c^m_{\nu,\eta}$. In Fig.~\ref{fig:coeff_ll_mm_spin_weights}(a)-(b) we display the resulting Fourier coefficients entering the superconducting order parameter. As seen, even though the onsite contribution dominates by one order of magnitude compared to the next leading contribution, there are significant longer-range couplings to neighboring unit cells.

As seen from Fig.~\ref{fig:gap_amplitude}, the momentum dependence of the gap is extreme at the BZ boundaries where the gap vanishes. In order to understand this property, we show in Fig.~\ref{fig:coeff_ll_mm_spin_weights}(c,d) the sublattice and spin-resolved weights on the $\alpha$-band along the momentum path $\Gamma-X-M-\Gamma$ [Fig.~\ref{fig:coeff_ll_mm_spin_weights}(c)] and along $\Gamma-Y-M'-\Gamma$ [Fig.~\ref{fig:coeff_ll_mm_spin_weights}(d)]. As seen, on the BZ boundaries the weights are spin and sublattice pure, but feature zero overlap between same-sublattice opposite-spin states, blocking any spin-singlet pairing on the boundaries. This is the main reason for the generated momentum-dependence of the superconducting gap in band space. We stress that this ``blocking effect" takes place only near the BZ edges and not throughout the entire BZ~\cite{Chakraborty2024}. The extent of the blocking mechanism is dependent on the magnitude of the Néel order parameter $N$; an increase of $N$  results in a more pronounced inward expansion of the blocking towards the BZ center. 
More specifically, the blocking originates from an interplay of the momentum dependence of the intra-sublattice hopping $t_{x,\mathbf{k}}$ and the anisotropic hopping term $t_{z,\mathbf{k}}$. In case of a nonsymmorphic altermagnet
as portrayed in Figs.~\ref{fig:gap_amplitude} and  \ref{fig:coeff_ll_mm_spin_weights}, $t_{z,\mathbf{k}}$ has symmetry-enforced nodes along the BZ boundary which coincide with the nodes of $t_{x,\mathbf{k}}$, such that the only breaking of degeneracy is due to the altermagnetic order. In the absence of SOC, the nonsymmorphic symmetry ensures these band degeneracies on nodal planes~\cite{Roig_Minimal_2024}. Hence, $l_{\downarrow,\mathbf{k}}$ and $m_{\uparrow,\mathbf{k}}$ are both zero on the BZ boundary which ultimately leads to the observed blocking effect.
We note further that for type IV altermagnets which feature no spin splitting in the entire BZ, the same arguments apply.
In case of symmorphic altermagnets, relevant e.g. for the Lieb lattice case, where $t_{z,\mathbf{k}} = t_4(\cos{k_x}-\cos{k_y})$ with nodes along the BZ diagonals, the nodes of $t_{x,\mathbf{k}}$ and $t_{z,\mathbf{k}}$ coincide only at the $M$-point. Hence, the strongest momentum dependence of the superconducting gap is observed at the corners of the BZ. Quantitatively, the extent of this effect again depends on the  Néel order parameter $N$. With increasing $N$ and above a critical value of the order of the altermagnetic hopping, the nodes can cover the entire BZ edges.

\subsection{NN interaction}
\begin{figure}
    \centering
    \includegraphics[width=\linewidth]{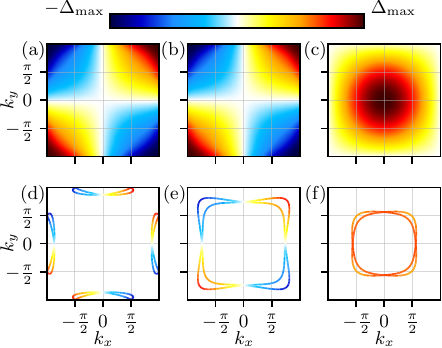}
    \caption{
    The superconducting order parameter stabilized by NN bare attraction in the regime of weak to intermediate altermagnetic spin splitting. The superconducting order parameter in the Brillouin zone for chemical potentials of (a) $0.3t$, (b) $0.6t$, and (c) $1.0t$. Panels (d)–(f) show the corresponding values of the superconducting order parameter projected onto the Fermi surface. The NN-interaction used and the self-consistent order parameters are for (a) $V^\mathrm{NN}=0.5t$ and $\Delta_{\mathrm{max}}=0.13t$, for (b) $V^\mathrm{NN}=0.6t$ and $\Delta_\mathrm{max}=0.23t$ and for (c) $V^\mathrm{NN}=0.9t$ and $\Delta_\mathrm{max}=0.17t$.}
    \label{fig:gap_amplitude_NN}
\end{figure}

Finally, we discuss the results for altermagnetic superconductivity stabilized by extended range attractions. In this case, for a normal nonmagnetic metal on the simple square lattice, the possible superconducting orders include extended $s$-wave, $d$-wave, and spin-triplet $p$-wave pairing~\cite{Roig2022}. In the present case, i.e. when the superconducting instability takes place inside an altermagnetic metal, there is again additional dressing from the sublattice-to-band transformation. In Fig.~\ref{fig:gap_amplitude_NN}, we show representative cases for the resulting gap structure for the same Fermi surfaces as above, but now with attractive NN pairing.
As seen, depending on the Fermi surface shape and the given amplitude of the pairing interaction, the preferred pairing symmetry is either a spin-singlet $d_{xy}$-wave or an extended $s$-wave state.
The edges of the BZ no longer block pairing since NN can pair e.g. A-sublattice spin-up with B-sublattice spin-down in agreement with the observation in  Fig.~\ref{fig:coeff_ll_mm_spin_weights}(c)-(d). The difference from the OS-case originates both from the different combinations of the eigenvectors $l_{\sigma,\mathbf{k}}$ and $m_{\sigma,\mathbf{k}}$ in the pairing kernel and the explicit momentum dependence of the bare NN-interaction. The latter contribution dominates the momentum dependence in the present cases. 

\begin{figure*}[t]
    \centering
    \includegraphics[width=\linewidth]{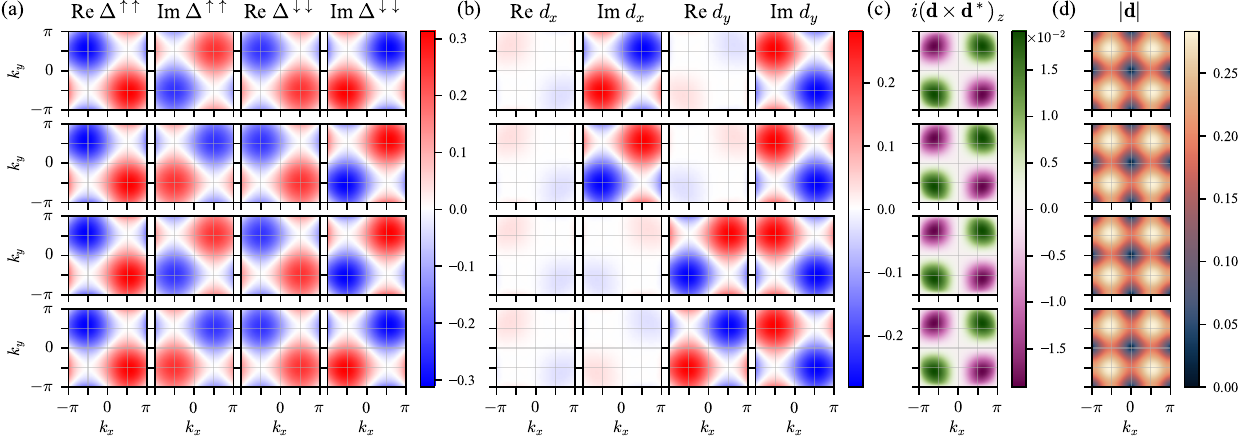}
    \caption{
    The superconducting order parameter stabilized by NNN bare attraction. Each horizontal row displays the momentum dependence of the ESP triplet states. All four states are degenerate. In the panels in (a) we show the real and imaginary parts of $\Delta^{\uparrow\uparrow}_{\mathbf{k}}$ and $\Delta^{\downarrow\downarrow}_{\mathbf{k}}$ where an overall phase from the random complex-valued initial guess has been removed, whereas panels in (b) display the associated d-vector components.  Panel (c) reveal the non-unitary nature of the ESP states. Finally, panel (d) show that all four states exhibit identical $|\bf d|$. Since $|\bf d|$ and $\mathbf{d}\times\mathbf{d}^*$ are identical for all four states, they exhibit the same spectroscopic gap and ground state energy. For all four self-consistent runs the parameters are  $V^\mathrm{NNN}=1t,N_z = 0.2t, \mu=0.6t$.}
    \label{fig:ESP}
\end{figure*}

The results shown in Fig.~\ref{fig:gap_amplitude_NN} are representative of the regime of small to intermediate amplitude of the spin splitting of the altermagnetic bands compared to the pairing amplitude. In the limit of large spin splitting compared to the superconducting gap scale, ESP triplet states become dominant. In this respect, earlier theoretical studies have discussed the four degenerate solutions $(p_x\pm i p_y) |\!\!\uparrow \uparrow \rangle \otimes (p_x\pm i p_y) |\!\!\downarrow \downarrow \rangle$, leading to a fully-gapped state for each spin sector~\cite{Zhu2023,Hong2025,Monkman,Heinsdorf_proxy}. The four different possibilities contain both helical and chiral phases and exhibit associated topological properties~\cite{Zhu2023}.

\subsection{NNN interaction}
In Fig.~\ref{fig:ESP} we show the results of ESP states obtained for the case of NNN attraction with sublattice dependency included for parameters $V^{\mathrm{NNN}}=1t,N_z=0.2t$ and $\mu=0.6t$. In that case, ESP triplet constitute the preferred superconducting ground state
\footnote{The triplet state would be stabilized for this set of parameters even in the absence of altermagnetic order. However, as we will demonstrate below, for $t_z \neq 0$ and $N_z \neq 0$, features unique to altermagnetism will be imprinted on the triplet state.}.
Following standard notation, we can parametrize the triplet state as $\Delta = i \sigma_y ({\bf{d}} \cdot \boldsymbol{\sigma})$, where ESP corresponds to $d_z=0$.
As seen in the four rows in panel (a) of Fig.~\ref{fig:ESP}, there are indeed four degenerate solutions (in the absence of SOC) in agreement with the previous studies.
Different from the earlier works, we find additional structure of the resulting $\bf{d}$-vector, as shown in the four rows in panel (b), yielding a non-unitary contribution $\bf{d}\times \bf{d}^*\neq 0 $ as shown in panel (c). This non-unitary contribution averages to zero over the entire BZ, and is a natural consequence of the altermagnetic spin-splitting.
Explicitly, let
\begin{equation}\label{eq:triplet_states}
    \Delta_\pm^{\uparrow \uparrow} \propto p_{x-y} \pm i \alpha p_{x+y}, \quad \Delta_\pm^{\downarrow \downarrow} \propto \alpha p_{x-y} \pm i p_{x+y},
\end{equation}
where the two $\pm$ are independent of each other, 
and $p_{x-y} ~=~ \sin k_x - \sin k_y$ and $p_{x+y} ~=~ \sin k_x + \sin k_y$ denote the odd-parity states formed by NNN pairing.
Since the Fermi surface of a specific spin breaks the $C_4$ rotational symmetry, $p_{x+y}$- and $p_{x-y}$-wave are not degenerate.
This results in an anisotropy where $p_{x-y}$ ($p_{x+y}$) is the dominant and $p_{x+y}$ ($p_{x-y}$) is the subdominant order parameter for spin up (down).
We parametrize this anisotropy in the superconducting state through the coefficient $\alpha < 1$.
This gives a non-unitary contribution 
\begin{equation}
    {\bf{d}}\times {\bf{d}}^* \propto (1-\alpha^2) (p_{x-y}^2 - p_{x+y}^2),
\end{equation}
when $\alpha \neq 1$, which vanishes when $k_x=0$ or $k_y=0$.
In the absence of the altermagnetic spin-splitting ($t_{z, \textbf{k}}=0$), $\alpha=1$ and hence gives zero non-unitary contribution.
We believe this property of the ESP states was not exposed in previous studies due to a restricted ansatz in the solution of the gap equations~\cite{Zhu2023,Hong2025,Monkman}.

Following the convention of Eq.~(\ref{eq:triplet_states})
the four degenerate states $\Delta_\pm^{\uparrow\uparrow}\otimes\Delta_\pm^{\downarrow\downarrow}$ shown in Fig.~\ref{fig:ESP}, from top to bottom carry the sign signatures $(+,-)$, $(-,+)$, $(+,+)$ and $(-,-)$.
The first two are frequently referred to as helical and the latter two as chiral $p$-wave states. The helical pairing states have Chern number $C=0$ while the two chiral states have $C= \pm 2$~\cite{Zhu2023,Monkman}.
Not all four states are invariant under the combined symmetry $C_4 \mathcal{T} $of the altermagnetic normal state. Only the first two, the helical states, preserve the symmetry of the normal state. 
Thus, the four states consist of two sets of double-component order parameters in agreement with Ref.~\cite{khodas2025tuningaltermagnetismstrain}, though in our case all four state are degenerate.
As seen from Fig.~\ref{fig:ESP}, the overall sign of the non-unitary component $\bf{d}\times \bf{d}^*$ is the same in all four degenerate ground states, and global reversal of the vector $\bf{d}\times \bf{d}^*$ is only achieved through sign change of the product $N_z t_z$ (i.e. reversal of Néel order $N_z\mapsto -N_z$ or $t_{z,\bf{k}} \mapsto -t_{z,\bf{k}}$) reflecting the spontaneous symmetry breaking at the transition to the magnetic state.
\begin{figure}[h!]
    \centering
    \includegraphics[width=\linewidth]{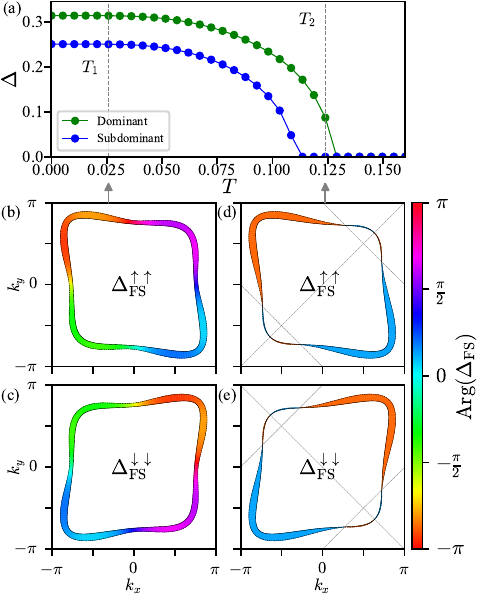}
    \caption{Temperature dependence and gap structure of the superconducting order parameters for the ESP state. Panel (a) displays the dominant (green) and subdominant (blue) superconducting order parameters as a function of temperature. Panels (b)-(e) show the superconducting order parameters on each spin Fermi surface for two specific temperatures: the fully gapped structure in (b) and (c) at $T_1=0.0256t$, and the nodal gap structure in (d) and (e) at $T_2 = 0.124t$. The contour width represents the magnitude of the superconducting order parameter and the color denotes its phase. The magnitude has been normalized to enhance visual clarity.
    The dashed lines in (d) and (e) mark the nodes of the superconducting order parameter.
    The parameters used are the same as those in Fig.~\ref{fig:ESP}.}
    \label{fig:high_temp}
\end{figure}

The chiral and helical states discussed above form at low temperatures, where both the dominant and subdominant $p$-wave components have developed in each spin sector.
However, the transition into the chiral or helical state as temperature decreases is not direct. Instead it consists of two distinct thermodynamic phase transitions.
The breaking of $C_4$ symmetry of each spin Fermi surface results in a dominant and subdominant order parameter component with strictly different critical temperatures.
At the highest critical temperature $T_c$, superconductivity develops only in the dominant component, that is $p_{x-y}$ for spin up and $p_{x+y}$ for spin down. 
Note that this phase is nodal, and corresponds to the case $\alpha=0$ in Eq.~(\ref{eq:triplet_states}). This phase also preserves a two-fold rotation symmetry. This can be seen in the $p_{x-y}$ momentum dependence of  $\Delta^{\uparrow\uparrow}$, which is invariant under a two-fold rotation ($C_{2,x-y}$) about the $(\hat{x}-\hat{y})/\sqrt{2}$ axis.
At some lower temperature $T'_c$, the subdominant component develops (i.e. $\alpha \neq 0$) and the chiral or helical state is entered. Note that in this state the previously mentioned two-fold rotation symmetry, $C_{2,x-y}$ is now broken, allowing the appearance of a second phase transition.
In contrast to the nodal phase at higher temperature, the chiral and helical states are fully gapped.
This is the microscopic mechanism which favors the helical and chiral states at low temperatures.
We demonstrate the existence of the two phases with a concrete example. 
Let us consider the same parameters used to find the helical and chiral states in Fig.~\ref{fig:ESP}, but now solve the gap equation self-consistently for a range of temperatures. The temperature dependence of the dominant and
483
subdominant order parameters are shown in Fig.~\ref{fig:high_temp}(a).
At low temperatures the fully gapped chiral state $(+,+)$ is formed, as seen in Figs.~\ref{fig:high_temp}(b) and \ref{fig:high_temp}(c).
As temperature increases, the subdominant component in each spin sector vanishes, entering the nodal phase ($p_{x-y}$ for spin up and $p_{x+y}$ for spin down) as seen in Figs.~\ref{fig:high_temp}(d) and \ref{fig:high_temp}(e). 
The difference between the two critical temperatures depends on the magnitude of the altermagnetic order parameter. For very large spin-splitting, the superconducting phase can be dominated by the nodal phase.

We stress that the presence of a double transition applies to the present case where the spin sectors are decoupled both in the normal state Hamiltonian and at the interaction level. For the more general case where the two sectors are coupled, the phase diagram will be modified.

\section{Discussion and Conclusions}\label{sec:conclusions}

Starting from minimal models that incorporate the sublattice degree of freedom for altermagnets, we have performed a self-consistent evaluation of the superconducting pairing in band space. The sublattice-to-band transformation imprints its structure on the superconducting gap parameter and provides significant momentum dependence for cases where the Fermi surface overlaps with the BZ edges. In the opposite case, i.e. when the Fermi surface is well contained inside the BZ, the one-band approximation can be a reasonable simplification. However, as shown in Ref.~\cite{Roig_Minimal_2024}, altermagnetic order becomes favorable when interband contributions are important. Thus, we expect quite generally that the Fermi surfaces of altermagnetic materials often will cross the BZ edges, and therefore a significant superconducting gap anisotropy is expected for nonsymmorphic altermagnetic superconductors. While we have focused on altermagnets, our results are also relevant for antiferromagnetic metals unstable to superconductivity. These two phases of matter, altermagnetism and antiferromagnetism, are smoothly connected in the minimal model by the parameter $t_{z,\mathbf{k}}$ in the absence of SOC. Antiferromagnetic metals that become superconducting will, in general, also exhibit inherent momentum-dependent gaps due to the sublattice degree of freedom~\cite{RomerAF}. 

In the cases where superconductivity is generated by longer-ranged attractions, the preferred superconducting ground state depends on an interesting interplay between Fermi surface shape, sublattice effects, and the ratio between altermagnetic spin-splitting and the amplitude of the superconducting gap. In the regime of significantly larger spin-splitting compared to the superconducting energy scale, ESP triplet superconductivity is favored. These $p$-wave triplet states can be of multi-component nature at low temperatures, featuring helical or chiral combinations. We have found that ESP triplet states exhibit a characteristic non-unitary property which is directly tied to the
the breaking of the $C_4$ rotation symmetry for a band of fixed spin species.
Experimentally, the predicted momentum-dependent gap structures can be assessed by spectroscopic probes including e.g. quasiparticle interference and photoemission measurements. The resulting gap anisotropy will also directly affect the density of states, and thereby the resulting temperature dependencies of e.g. the specific heat and the penetration depth. The ESP triplet pair states exhibit additional characteristic signatures that can be tested experimentally. For example, these composite ESP condensates can generate spin-polarized persistent currents, and support pure spin superflow composed of counter-propagating spin-up and spin-down supercurrents~\cite{Monkman}. Additionally, the ESP phases are topological and feature associated topological surface states~\cite{Zhu2023,Monkman}. Depending on whether the preferred ground state is helical or chiral, characteristic currents should form near defects and other sample inhomogeneities~\cite{Roig2022,Clara,Andersen2024}. Associated magnetic fields can possibly be detected by muon spin rotation or nitrogen-vacancy centers~\cite{Andersen2024,Jyoti}. 

Although an altermagnetic superconductor is still awaiting definite experimental discovery at present, several relevant materials have been recently identified. For example, recent research has revealed that thin films of $\mathrm{RuO}_2$ become superconducting under strain~ \cite{Uchida_Superconductivity_2020,Ruf2021,Occhialini,Wadehra2025}. In addition, other altermagnetic metals, such as KV$_2$Se$_2$O, CrSb, MnTe, or CoNb$_4$Sb$_8$ may also exhibit superconductivity at low temperatures when exposed to pressure, doping or dimensional reduction~\cite{Krempasky_Altermagnetic_2024,Reimers2024,Jiang2025,Regmi2025,Federico2025}. Certainly, the rapid current development of altermagnetic metals provides a promising avenue for future theoretical and experimental studies of the fascinating interplay between altermagnetism and superconductivity.

\begin{acknowledgments}
We acknowledge useful discussions with R. M. Fernandes,
M. Franz, and M. Khodas. J.G. acknowledges support from the Independent Research Fund Denmark, Grant No. 3103-00008B. M.B was supported by a research grant (VIL69220) from VILLUM FONDEN. B.M.A. acknowledges support from the Independent Research Fund Denmark Grant No. 5241-00007B. A.K. acknowledges support by the Danish National Committee for Research Infrastructure (NUFI) through the ESS-Lighthouse Q-MAT. 
D.F.A was supported by National Science Foundation Grant No. DMREF 2323857. Work at UWM  was also supported by a grant from the Simons Foundation (SFI-MPS-NFS-00006741-02, M.R.).
\end{acknowledgments}


%
\end{document}